\documentclass[aoas,MSNbibl,nameyear,seceqn,dvips]{arximspdf}
\usepackage{dcolumn}

\doi{10.1214/12-AOAS560} 
\volume{6}
\issue{4}
\pubyear{2012}
\firstpage{1641}
\lastpage{1663}

\makeatletter

\newcolumntype{d}[1]{D{.}{.}{#1}}

\makeatother

\begin{document}
\begin{frontmatter}

\title{Assessing transient carryover effects in recurrent event processes,
with application to chronic health~conditions}
\runtitle{Carryover effects in recurrent event processes}

\begin{aug}
\author[A]{\fnms{Candemir} \snm{\c{C}\.{i}\u{g}\c{s}ar}\corref{}\thanksref{t1}\ead[label=e1]{candemir.cigsar@wchospital.ca}}
\and
\author[B]{\fnms{Jerald F.} \snm{Lawless}\thanksref{t2}\ead[label=e2]{jlawless@math.uwaterloo.ca}}
\runauthor{C. \c{C}i\u{g}\c{s}ar and J. F. Lawless}
\affiliation{Women's College Hospital and
Princess Margaret Hospital, and~University~of~Waterloo}
\address[A]{Women's College Research Institute \\
Women's College Hospital \\
University of Toronto\\
790 Bay Street, 7th Floor\\
Toronto, Ontario M5G 1N8 \\
Canada\\
\printead{e1}} 
\address[B]{Department of Statistics\\
\quad and Actuarial Science\\
University of Waterloo\\
200 University Avenue West\\
Waterloo, Ontario N2L 3G1 \\
Canada\\
\printead{e2}}
\end{aug}

\thankstext{t1}{Supported by Cancer Care Ontario and the Ontario
Institute for Cancer Research (through funding provided by the Ministry
of Health and Long-Term Care and the Ministry of Research and Innovation
of the Government of Ontario).}

\thankstext{t2}{Supported by the Natural Sciences and Engineering
Research Council
of Canada.}

\received{\smonth{7} \syear{2011}}
\revised{\smonth{4} \syear{2012}}

%
\begin{abstract}
In some settings involving recurrent events, the occurrence of one
event may produce a temporary increase in the event intensity; we refer
to this phenomenon as a transient carryover effect. This paper provides
models and tests for carryover effect. Motivation for our work comes
from events associated with chronic health conditions, and we consider
two studies involving asthma attacks in children in some detail. We
consider how carryover effects can be modeled and assessed, and note
some difficulties in the context of heterogeneous groups of
individuals. We give a simple intuitive test for no carryover effect
and examine its properties. In addition, we demonstrate the need for
detailed modeling in trying to deconstruct the dynamics of recurrent
events.
\end{abstract}

%

\end{frontmatter}

\section{Introduction}\label{sec1}

Recurrent events experienced by individuals, units or systems occur in
many fields [\citet{CooLaw07}]. For example, repeated failures
can occur for equipment or for software systems [\citet{AscFei84},
\citet{Bak01}, \citet{Lin06}]. In medical contexts,
individuals may experience multiple episodes of hospitalization,
recurrent infections or children may suffer repeated attacks of asthma
[\citet{Ducetal03}]. Models for recurrent events are discussed in
books such as \citet{CoxIsh80} and \citet{DalVer03}.
\citet{CoxLew66}, \citet{Kar91} and \citet{CooLaw07} discuss
related methods of analysis.

In certain settings an event intensity is temporarily increased (or in
some cases, decreased) after some condition or event occurs. Such
transient effects may be due to factors that are either external or
internal to the individuals or systems in question. Transient effects
due to external factors have received considerable recent attention.
For example, \citet{FarWhi06} and \citet{FarHoc10} have
examined potential adverse health effects following
administration of the mumps, measles and rubella (MMR) vaccine to
children. \citet{FarWhiHoc09} consider adverse
effects related to drug treatments. Our focus in this paper is on
internal factors. These are not usually observable, and evidence for
their existence is sought by examining whether event intensities are
temporarily increased soon after an event occurs. We call such effects
carryover effects, which is also a term used to describe transient
effects due to external factors such as vaccinations or residual
effects of drugs [\citet{CooLaw07}, Section 3.8.2]. This
phenomenon has also often been discussed for hardware or software
systems, where repairs or modifications undertaken to deal with a
failure may not resolve the problem or may even create new problems
[e.g., Baker (\citeyear{Bak96}, \citeyear{Bak01}), \citet{Pen06}].

The motivation for the present paper is from attempts to identify
potential carryover effects related to events occurring in subjects
with chronic medical conditions. Such effects are inherently difficult
to assess because of complex factors that may influence event
occurrence. These include unobservable covariates that can produce wide
heterogeneity in event rates across individuals and the presence of
temporal trends that may be related to the age of a process or to
external factors such as seasonal effects. In addition, clinical events
are often related to unobservable processes concerning a person's
health and fluctuations in such processes can produce clustering of
events. This paper is motivated specifically by studies of adverse
events in children. Two studies that we consider here involve
randomized treatment trials for the prevention of asthma attacks; a
third study that will be discussed more briefly later in the paper
involves failures associated with shunts which are used to drain excess
cerebrospinal fluid in children with hydrocephalus [\citet{Tuletal00}].

In the first asthma prevention trial, infants who were considered at
high risk for asthma were randomized at 6 months of age to receive
either a placebo or drug treatment [\citet{Ducetal03}]. They then
were followed for 18 months, and occurrences of any asthma attacks
(according to specified symptoms) were recorded. In addition to the
assessment of any drug effect, other points of interest are the
evolution of the asthma recurrent event rate over time and how the
occurrence of an event influences the event rate [\citet
{Ducetal03}, page 356]. In the second study [\citet{Veretal03}], children
aged 4--11 years were randomized to receive either 200 or 400 $\mu$g
per day of fluticasone propionate (FP) for the prevent of asthma
exacerbations. The original protocol called for 3 months of follow-up
per child, but this was later amended to 12 months. Most of the
exacerbations in question were defined as ``moderate;'' these were
defined as occurring if a child experienced a period of two consecutive
days on which either (i) their morning percentage predicted
expiratory flow (PEF, a measure of lung function) fell more than 20\%
below their baseline value measured at randomization, or (ii) they
had an increase in inhaler ($\beta_{2}$-agonist) usage.\looseness=1

In each of these studies we will examine whether there is an indication
that individuals are temporarily at a higher risk of a new event
(exacerbation) following the resolution of a previous exacerbation.
Insights into this can affect strategies for the prevention and
treatment of exacerbations. As an illustration we show a simple
synopsis of data from the first asthma trial, in which 119 children
were randomized to the placebo control group and 113 were randomized to
the treatment group. The total numbers of asthma attacks were 483
(control group) and 336 (treatment group). The total observed and
expected (calculated under a hypothesis of no carryover effect, as
described in Section~\ref{sec5.1}) number of attacks which occurred
within two
weeks of the preceding attack are as follows:
\begin{eqnarray*}
\mbox{\textit{Control group}:\quad} \mbox{\textit{Observed}} &=&
121,\qquad
\mbox{\textit{Expected}}= 80.3, \\
\mbox{\textit{Treatment group}:\quad} \mbox{\textit{Observed}} &=&
76,\qquad
\mbox{\textit{Expected}} = 40.5.
\end{eqnarray*}
The data show an excessive number of events soon after the preceding event.

The presence of a carryover effect can be assessed fairly readily in
single systems which experience large numbers of events [e.g.,
\citet{Bak01}]. However, in medical contexts we typically have a
large number
of individuals, each with a small number of events. The purpose of this
paper is to discuss models through which the presence of a carryover
effect can be assessed in settings involving multiple heterogeneous
individuals, as seen in the preceding studies. We make three novel
contributions. First, we show that internal carryover effects can be
difficult to distinguish from subject heterogeneity in settings where
the average number of events per subject is fairly small. Second, we
show that the data often have limited information about the duration of
an effect, so reliance on background information is crucial. Finally,
we provide tests for no carryover effect which are simple to interpret
and reasonably robust.

In Section~\ref{sec2} we consider models for transient carryover
effects, discuss their connection to the concept of event clustering,
and show how heterogeneity makes the assessment of transient effects
more difficult. Section~\ref{sec3} considers some simple tests and
Section~\ref{sec4} presents simulation results on their properties.
Section~\ref{sec5} examines the studies on asthma in infants. Section
\ref{sec6} contains concluding remarks and discusses a study of
cerebrospinal fluid shunt failures in pediatric patients. In the
interests of exposition, some technical derivations are placed in the
\hyperref[app]{Appendix}.

\section{Models for carryover effects}\label{sec2}

We use standard notation for recurrent events. We assume that an
individual process is observed over time interval $[0,\tau]$, and let
$N (t)$ denote the number of events in $(0, t]$. The history of events
over $[0, t)$ is denoted by $\mathcal{H} (t)$ and the event intensity
function [\citet{CooLaw07}, page 10] is given by
%
%
\begin{equation}\label{e2.1}
\lambda(t | \mathcal{H} (t)) = \lim_{\Delta t \downarrow0}
\frac{\Pr
\{
N (t + \Delta t^{-}) - N (t^{-}) = 1 | \mathcal{H} (t)
\}}
{\Delta t}.
\end{equation}
The intensity fully specifies continuous time processes where at most
one event can occur at a given time. The times of events are denoted
$T_{1} < T_{2} < \cdots\,$, and $B (t) = t - T_{N(t^{-})}$ is the elapsed
time since the most recent event prior to $t$. Familiar models include
Poisson processes, where $\lambda(t | \mathcal{H} (t)) = \rho(t)$ for
some function $\rho$, and renewal processes, where $\lambda(t |
\mathcal{H} (t)) = h (B (t))$ for some function $h$.

Carryover effects can be modeled in a number of ways. A model that is
very useful when events may display time trends is a modulated Poisson
process. In this case, (\ref{e2.1}) takes the form
%
%
\begin{equation}\label{e2.2}\label{Eqmodel1}
\lambda(t | \mathcal{H} (t)) =
\rho_{0} (t) \exp(\beta^{\prime} Z (t)),
\end{equation}
where $Z (t)$ is a $q \times1$ vector of time-varying covariates that
is allowed to contain functions of the event history $\mathcal{H} (t)$
as well as external covariates. More specifically, we can consider
models for which $Z (t)$ includes terms that are zero except for a
limited time period following the occurrence of an event; such terms
specify the carryover effects. A simple but very useful model is one
where $Z (t)$ in (\ref{e2.2}) takes the form
%
%
\begin{equation}\label{e2.3}\label{Eqcoeffect}
Z (t) = I \bigl( N (t^{-}) > 0 \bigr) I \bigl(B (t) \leq\Delta\bigr),
\end{equation}
where $\Delta> 0$ is a specified value. In that case the intensity
function following an event temporarily changes from $\rho_{0} (t)$ to
$e^{\beta} \rho_{0} (t)$. Tests of the null hypothesis $H_{0}\dvtx\beta=
0$, developed below, provide simple and intuitive tests of no carryover effect.

Other similar models with carryover effects can also\vspace*{1pt} be specified. For
example, a model (\ref{e2.2}) with $Z (t) = I (N (t^{-}) > 0) \exp(-
\gamma B
(t))$ or an additive linear self-exciting process [\citet
{CoxIsh80}, Section 3.3;
\citet{Oga83}] with $\lambda(t | \mathcal{H}(t)) =
\rho_{0} (t) + \beta\sum^{N (t^{-})}_{j = 1} e^{- \gamma(t - t_{j})}$
also produces transient effects following events, while allowing
possible time trends as in (\ref{e2.2}). Such models are more difficult to
handle than (\ref{e2.2}) and (\ref{e2.3}), and do not impose a time limit
on the
duration of an effect, but have been found useful in areas such as
seismology [\citet{Oga83}].

There is a close connection between what we term carryover effects and
cluster processes [\citet{CoxIsh80}, Section 3.4]. In a cluster
process the events occur in clusters, or groups of events that are
close together in time. Carryover effects in essence produce a type of
clustering, and models such as (\ref{e2.3}) or the linear self-exciting
process can be viewed as cluster processes in which each new event
produces a subprocess going forward in time, with a decreasing rate
function [e.g., \citet{CoxIsh80}, pages 69, 77]. On the basis of
observed events alone, it is impossible to say what produces observed
clustering, and we must rely on context-specific background to suggest
plausible mechanisms. We view internal carryover effects as arising
when a ``remedy'' for an adverse effect is unsuccessful or partially
successful, and consider models that facilitate interpretation within
that framework. Many models for cluster processes are harder to handle
[e.g., \citet{CoxIsh80}, Section 3.4; \citet{XieSunNau09}],
especially when the rate of events is not stationary, and there is
heterogeneity. Standard clustering models do not address these points.
Our models are straightforward to handle and provide insight, but as
always, models should be checked, and other approaches may be needed
in some situations. We note as well that although we focus on the
case where the intensity temporarily increases following an event, in
some contexts it could decrease, with $\beta$ in (\ref{e2.2}) being negative
in that case.

Another way to consider carryover effects is through the distribution
of gap times $W_{j} = T_{j} - T_{j-1}$ (with $T_{0} = 0$) between
successive events. Gap time models [\citet{CooLaw07}, Chapter 4]
are particularly useful in settings where an adverse event results in
some corrective action which ideally returns an individual to a ``good
as new'' state [e.g., \citet{Pen06}]. Gap time models in which the
times between successive events have distributions with substantial
mass near zero could be considered as suggesting a carryover effect
[e.g., see \citet{Bak01}, \citet{Lin06}, \citet{Pen06}]. They
contain more parameters and are more difficult to handle than
(\ref{e2.2}) and (\ref{e2.3}), and do not accommodate calendar time trends
as readily,
but are often useful. In the special stationary case where $\rho_0 (t)$
in (\ref{e2.2}) is a constant $\alpha$, the model with $Z(t)$ given by
(\ref{e2.3})
is a delayed renewal process where the times $W_{j}$ ($j=2, 3,
\ldots$) between successive events are independent random variables
with a hazard function of the form $h(w) = \alpha e^{\beta} I(w \leq
\Delta) + \alpha I(w > \Delta)$.

In applications involving multiple systems or individuals,
heterogeneity is often apparent [e.g., \citet{Law87},
\citet{Bak01},
\citet{Lin06}, \citet{CooLaw07}, Section 3.5]. For example,
individual processes may be (approximately) Poisson, but their rate
functions may vary. Such variation is typically due to unmeasured
differences in the individuals or the environment in which the
processes operate. It is imperative to consider the possibility of
heterogeneity because, as we show below, it can create an appearance of
a carryover effect when no such exists.

The simplest and most useful extension of modulated Poisson process
models (\ref{e2.2}) to include heterogeneity is where independent
processes $i
= 1, \ldots, m$ have rate functions
%
%
\begin{equation}\label{e2.4}
\rho_{i} (t | \mathcal{H}_{i}(t)) =
\alpha_{i} \rho_{0} (t) \exp( \beta Z_{i} (t) ),
\end{equation}
where $\alpha_{1}, \ldots, \alpha_{m}$ are positive-valued variables.
Models for which the $\alpha_{i}$ are fixed parameters can be
problematic because the $\alpha_{i}$ cannot be estimated consistently.
An alternative is to assume the $\alpha_{i}$ are independent and
identically distributed random effects with some distribution function
$G (\alpha; \phi)$, where $\phi$ is a vector of parameters [\citet
{CooLaw07}, Section 3.5], and we consider this for most analyses.

We now show why heterogeneity that is not taken into account can
misleadingly suggest a carryover effect. Suppose for illustration that
the model (\ref{e2.4}) with $\beta= 0$ and $\alpha_{i}$ following a gamma
distribution describes a situation. Without loss of generality, we take
the $\alpha_{i}$ to have mean 1 and variance $\phi$, and then
[\citet{CooLaw07}, page 79] we find that the intensity function
for the
process with the unobservable $\alpha_{i}$ integrated out is
%
%
\begin{eqnarray}\label{e2.5}
\lambda_{i}(t | \mathcal{H}_{i} (t))
&=&
E (
\alpha_{i} | \mathcal{H}_{i}(t)
)
\rho_{0}(t) \nonumber\\[-8pt]\\[-8pt]
&=&
\biggl\{
\frac{\phi^{-1} + N_{i}(t^{-})}
{\phi^{-1} + \int_{0}^{t} \rho_{0}(u) \,du}
\biggr\}
\rho_{0}(t).\nonumber
\end{eqnarray}
Note that when an event occurs, the numerator term in brackets in (\ref{e2.5})
increases by one, thus increasing the intensity. As $t$ increases up to
the next event, the denominator in brackets increases, so the overall
effect is that the intensity increases immediately after an event
occurs and then decreases. This is the type of behavior we associate
with a carryover effect. The larger the degree of heterogeneity across
the individuals (i.e., the larger $\phi$ is), the larger is the
increase following an event. As $t$ becomes arbitrarily large, the term
in brackets converges in probability to $\alpha_{i}$ so the appearance
of a carryover effect is mainly in the earlier events. However, failure
to incorporate heterogeneity in models can produce spurious evidence of
an effect. To demonstrate, we ran a small simulation by generating
1000 realizations of a random effect model without carryover effects;
we used model (\ref{e2.4}) where the $\alpha_{i}$ have a gamma distribution
with mean 1 and variance $\phi$ and parameters $\rho_{0}(t) = \gamma$
and $\beta= 0$ (no carryover effect). We considered eight scenarios
with various combinations of $\gamma$, $\phi$ and $m$ ($\gamma= 2$, 5,
$\phi= 0.2$, 0.5 and $m = 100$, 500). Observation periods were $(0,
\tau_{i})$ and the $\tau_i$ times were generated from a uniform
distribution over $(0.8, 1.2)$. For $\tau_{i} = 1$ the expected number
of events per individual is 2 or 5 when $\gamma= 2$ or 5,
respectively. For each sample we obtained the maximum likelihood
estimates of parameters and their standard errors in the carryover
effect model (\ref{e2.2}) with (\ref{e2.3}), without incorporating
heterogeneity. We
found that $\hat{\beta}$ was positively biased across the 1000
simulations for each scenario, with mean to standard deviation ratios
varying from 0.7 to over 10. Correspondingly, tests of the null
hypothesis $H_{0}\dvtx\beta= 0$ incorrectly reject $H_{0}$ with high
probability. Using the same data sets, we also fitted the carryover
effect model with random effects (\ref{e2.4}) with $\rho_{0}(t) = \gamma
$, and
in this case the means of $\hat{\beta}$ were close to zero for all
scenarios. The results can be found in the supplementary material
[\citet{CgsLaw}].

We remark also that heterogeneity is to some extent confounded with a
carryover effect even with a proper model specification. With (\ref{e2.4}),
for example, and gamma distributed $\alpha_i$, the intensity function is
\[
\lambda_{i}(t | \mathcal{H}_{i}(t)) =
\biggl\{
\frac{\phi^{-1}+N_{i}(t^{-})}
{\phi^{-1}+\int_{0}^{t}\rho_{0}(u) e^{\beta Z_{i}(u)} \,du}
\biggr\}
\rho_{0}(t)
e^{\beta Z_{i}(t)}.
\]
As $t$ becomes large the term in brackets once again converges in
probability to $\alpha_{i}$, so a carryover effect represented by
$\beta\neq0$ can be readily assessed. When $t$ and $N_{i}(t^{-})$ are
small, however, the carryover effect and the expression in brackets can
both produce substantial temporary increases in the event intensity. In
many of the applications we consider, there are many individuals but
relatively few events for most individuals, and therefore a process of
careful modeling and model-checking is warranted. Next, we consider
some tests of no carryover effect based on (\ref{e2.4}). These are reasonably
robust and have a simple interpretation in terms of the observed data.

\section{Tests based on Poisson processes}\label{sec3}

We consider tests of no carryover effect based on the Poisson model
(\ref{e2.4}), and testing that $\beta= 0$. This can be done either
using a
parametric model for $\rho_{0}(t)$ or by using a nonparametric
specification, in which case (\ref{e2.4}) is a modulated Andersen--Gill model
with frailty [\citet{CooLaw07}, page 81]. We describe the
parametric setting in detail, so as to show the intuitive form of the
test statistics, and then discuss the semiparametric case. The tests
use a specified value for $\Delta$ in (\ref{e2.3}) and (\ref{e2.4}).
This is
consistent with common practice and the resulting tests have the nice
form of a difference between observed and expected numbers of events in
the window of length $\Delta$ following an event. However, we later
consider the effects of misspecifying $\Delta$ and in Section \ref
{sec5} we
consider estimation of $\Delta$.

\subsection{Fixed effects model}\label{sec3.1}

We consider first the fixed effects model (\ref{e2.4}), for which the
$\alpha
_{i}$ are treated as unknown parameters. This can be useful when the
number of individual processes $m$ is small but there are many events
per process. The follow-up (censoring) times $\tau_{i}$ throughout the
paper are assumed to be stopping times [\citet{CooLaw07}, page
48]. The follow-up times are therefore allowed to be random and to
depend on previous event history. In this case, data on $m$ independent
processes give the log likelihood function
%
%
\begin{equation}\label{e3.1}\label{Eqfixloglike}
\ell(\alpha, \gamma, \beta) = \sum^{m}_{i = 1} \Biggl\{ n_{i} \log\alpha
_{i} + \sum^{n_{i}}_{j = 1} [ \log\rho_{0} (t_{i j}; \gamma) + \beta
Z_{i} (t_{i j}) ] - \alpha_{i} R_{i} (\gamma, \beta)\Biggr\},\hspace*{-30pt}
\end{equation}
where $\alpha= (\alpha_{1}, \ldots, \alpha_{m})^{\prime}$ and
%
%
\begin{equation}\label{e3.2}\label{Eqrint}
R_{i} (\gamma, \beta) =
\int^{\tau_{i}}_{0} \rho_{0} (t; \gamma) e^{\beta Z_{i}(t)} \,dt.
\end{equation}
For given $\gamma$ and $\beta$, (\ref{e3.1}) is maximized by $\tilde
{\alpha
}_{i} (\gamma, \beta) = n_{i} / R_{i} (\gamma, \beta)$, and
substitution of this into (\ref{e3.1}) gives the profile log likelihood for
$\gamma$ and $\beta$ as a constant plus
%
%
\begin{equation}\label{e3.3}
\ell_{p} (\gamma, \beta) = \sum^{m}_{i = 1} \Biggl\{ \sum^{n_{i}}_{j = 1} [
\log\rho_{0} (t_{i j}; \gamma) + \beta Z_{i} (t_{i j}) ] - n_{i} \log
R_{i} (\gamma, \beta) \Biggr\}.
\end{equation}
A likelihood ratio test of $H_{0}\dvtx\beta= 0$ requires estimates $\hat
{\gamma}, \hat{\beta}$ that maximize (\ref{e3.3}) and the estimate
$\tilde
{\gamma}$ that maximizes $\ell_{p} (\gamma, 0)$; the estimates are
found easily by general optimization software.

A score test can be based on $U_{\beta} (\tilde{\gamma}, 0)$, where
$U_{\beta} (\gamma, \beta) = \partial\ell_{p} (\gamma, \beta) /
\partial\beta$. The standardized score statistic
\begin{eqnarray*}
U_{\beta}(\tilde{\gamma}, 0) &=&
\sum_{i=1}^{m}
\Biggl\{
\sum_{j=1}^{n_{i}} Z_{i}(t_{ij}) -
\frac{ n_{i} \int_{0}^{\tau_{i}} Z_{i}(t) \rho_{0} (t; \tilde{\gamma})
\,dt}
{\int_{0}^{\tau_{i}} \rho_{0} (t; \tilde{\gamma}) \,dt}
\Biggr\}\\
&=& \operatorname{Obs} (\Delta) - \operatorname{Exp} (\Delta),
\end{eqnarray*}
where $\operatorname{Obs}(\Delta) = \sum_{i=1}^{m} \sum_{j=1}^{n_{i}} Z_{i}(t_{ij})$
is the observed number of events that occur within time $\Delta$ of a
preceding event, and $\operatorname{Exp} (\Delta)$ is an estimate of the expected
number of such occurrences under the hypothesis of no carryover effect.
For the simple case of a homogeneous Poisson process, $\rho_{0}(t;
\gamma)$ is one, and we find $\operatorname{Exp} (\Delta) = (n_{i}/ \tau_{i}) \sum
_{j=2}^{n_{i}+1} \min(w_{ij}, \Delta)$, where $w_{ij} = t_{ij} - t_{i,
j-1}$ ($j=1, \ldots, n_{i}$) and $w_{i,n_{i}+1} = \tau_{i} - t_{i
n_{i}}$. The form ``observed minus expected'' for $U_{\beta} (\tilde
{\gamma}, 0)$ is easily understood and useful. The standardized form of
$U_{\beta} (\tilde{\gamma}, 0)$ is [\citet{Pen98}]
%
%
\begin{equation}\label{e3.4}\label{Eqscorestat31}
S = \frac{ U_{\beta}(\tilde{\gamma}, 0)} { \widehat{\operatorname{Var}}
[ U_{\beta} (\tilde{\gamma}, 0) ]^{1/2} },
\end{equation}
where\vspace*{1pt} $\widehat{\operatorname{Var}} [ U_{\beta} (\tilde{\gamma}, 0) ] = \tilde
{I}_{\beta\beta} - \tilde{I}_{\gamma\beta} \tilde{I}_{\gamma\gamma
}^{-1} \tilde{I}_{\beta\gamma}$ is obtained from the observed
information matrix for $\beta$ and $\gamma$ based on (\ref{e3.3}), evaluated
at $(\gamma, \beta)=(\tilde{\gamma}, 0)$.

A problem with $S$, and with the likelihood ratio statistic, is that if
$m \rightarrow\infty$ but the $\tau_{i}$ are fixed, the limiting
distributions are not standard normal and $\chi^{2}_{(1)}$,
respectively, due to the fact that the $\alpha_{i}$ are not estimated
consistently. The normal and $\chi^{2}$ approximations may be adequate
in cases where $m$ is not too large and the numbers of events per
process are fairly large, but simulations in Section~\ref{sec4} show
they are
inadequate in settings like those in Section~\ref{sec5}. However, we
can use a
simulation (parametric bootstrap) approach to get $p$-values. Under
$H_0$, the event times $T_{i1}, \ldots, T_{in_{i}}$, given
$N_{i}(\tau_{i}) = n_{i}$, are the order statistics for a random sample
of size $n_{i}$ from the truncated distribution with density function
[\citet{CoxLew66}, Section 3.3]
\[
f_{i}(t; \gamma) =
\frac{\rho_0 (t;\gamma)}
{\int_{0}^{\tau_{i}} \rho_{0}(s; \gamma) \,ds},\qquad
0 \leq t \leq\tau_i.
\]
Thus, we can generate random samples from each $f_{i}(t; \tilde{\gamma
})$, $i=1, \ldots, m$, and use these to obtain values of the test
statistic in question. For the HPP case, $f_{i}(t; \gamma)$ is the
uniform distribution on $[0, \tau_i]$. It should be noted that
$p$-values obtained from this approach are conditional on the observed
values $n_1, \ldots, n_m$ and so are not strictly comparable to the
unconditional $p$-values provided by a normal or $\chi^{2}$
approximation, or to $p$-values for the random effects model in
Section~\ref{sec3.2}.

\subsection{Random effects model}\label{sec3.2}

Random effects models employ a distribution for the $\alpha_{i}$ in
(\ref{e2.4}), which are assumed independent. We assume for discussion
that the
$\alpha_i$ have a gamma distribution with mean 1 and variance $\phi$,
which is a widely used model; similar developments can be given for
other distributions. In this case the log likelihood function is
[\citet{CooLaw07}, Section 3.5.3]
%
%
\begin{eqnarray}\label{e3.5}\label{Earandloglike}
\ell(\gamma, \beta, \phi) & = &
\sum^{m}_{i = 1}
\Biggl\{ \sum^{n_{i}}_{j = 1}
[
\log\rho_{0} (t_{i j }; \gamma) + \beta Z_{i} (t_{i j})
]\nonumber\\[-1.5pt]
&&\hspace*{19.7pt}{} +
\log\Gamma
( n_{i} + \phi^{- 1} ) - \log\Gamma(\phi^{- 1})
\\[-1.5pt]
&&\hspace*{19.7pt}{}
+ n_{i} \log\phi- (n_{i} + \phi^{- 1})
\log[ 1 + \phi R_{i}(\gamma, \beta) ]
\Biggr\}.\nonumber
\end{eqnarray}
Likelihood ratio tests of $H_{0}\dvtx\beta= 0$ require maximum likelihood
estimates $\hat{\gamma}, \hat{\beta}, \hat{\phi}$ and $\tilde{\gamma},
\tilde{\phi}$ (when $\beta= 0$); these are readily obtained with
general optimization software. The $R_{i} (\gamma, \beta)$ in (\ref
{e3.5}) are
as defined in (\ref{e3.2}).

Score tests of $\beta= 0$ require only $\tilde{\gamma}$ and $\tilde
{\phi}$. Appendix~\ref{appB} gives the score statistic
%
%
\begin{equation}\label{e3.6}\label{Eqrandscoretest}
S =
U_{\beta} (\tilde{\gamma}, 0, \tilde{\phi})
/
\widehat{\operatorname{Var}} [ U_{\beta} (\tilde{\gamma}, 0, \tilde{\phi}) ]^{1/2}
\end{equation}
corresponding to (\ref{e3.4}). It is instructive to consider the
numerators of
(\ref{e3.4}) and (\ref{e3.6}); the numerator of (\ref{e3.6}) is (see
Appendix~\ref{appB})
%
%
\begin{equation}\label{e3.7}\label{Eqrandscore}
U_{\beta} (\tilde{\gamma}, 0, \tilde{\phi}) =
\operatorname{Obs} (\Delta) - \sum^{m}_{i = 1}
\frac{(1+n_{i} \tilde{\phi})
\int_{0}^{\tau_{i}} Z_{i}(t) \rho_{0}(t; \tilde{\gamma}) \,dt}
{1+ \tilde{\phi} \int_{0}^{\tau_{i}} \rho_{0}(t; \tilde{\gamma}) \,dt}.
\end{equation}
Equation (\ref{e3.7}) differs from the numerator of (\ref{e3.4}) in the
calculation
of the second term, $\operatorname{Exp}(\Delta)$. The fixed effects case (\ref{e3.4})
corresponds to the limit of (\ref{e3.7}) as the estimated
variance\vadjust{\goodbreak}
$\tilde{\phi
}$ of the $\alpha_{i}$ becomes arbitrarily large. Assuming that the
gamma distribution for the $\alpha_{i}$ is correct, the statistic $S$
in (\ref{e3.6}) is asymptotically $N(0, 1)$ as $m \rightarrow\infty$, unlike
the fixed effects statistic. In Section~\ref{sec4} we examine the
adequacy of
the normal approximation in practical settings. In situations where it
is inadequate we can use simulation (parametric bootstrap) to obtain
$p$-values. In addition, the gamma distribution will never be exactly
correct in practice, so we consider the performance of (\ref{e3.6}) under
departures from the gamma in Section~\ref{sec4}.

The Andersen--Gill model with random effects $\alpha_i$ [\citet
{CooLaw07}, page 81] can also be used. This model places no
parametric restrictions on $\rho_{0}(t)$ in (\ref{e2.4}). The R/S-Plus
function \texttt{coxph} with the frailty option implements this, but
some work is needed to extract \textit{Observed}--\textit{Expected} components
analogous to~(\ref{e3.7}); see Appendix~\ref{appB}.

\subsection{\texorpdfstring{Power of tests and choice of $\Delta$}{Power of tests and choice of Delta}}\label{sec3.3}

The tests of no carryover effect in the preceding section are based on
a specified value of $\Delta$ and a family of alternative hypotheses,
but are robust in the sense that the tests of the null Poisson
processes are, under some conditions, consistent against carryover
alternatives that are not in the family represented by (\ref{e2.2}) and
(\ref{e2.3}).
That is, as $m \rightarrow\infty$, the probability $H_{0}$ is rejected
approaches one under the alternative. We illustrate this property via
simulation in Section~\ref{sec4}, where we show that the tests in
Sections
\ref{sec3.1}
and~\ref{sec3.2} retain good power when $\Delta$ is misspecified and
when the
random effects distribution (Section~\ref{sec3.2}) is misspecified. Simulation
results (not shown) also indicate the tests retain power against
alternatives where the true form of the process intensity is additive
[$\lambda_{0} (t; \gamma)+ \beta Z_{i}(t)$] rather than multiplicative.
The model (\ref{e2.4}) should, however, be checked for consistency with the
data; ways to do this are discussed by \citet{CooLaw07},
Chapters 3 and 5. Carryover effects can also be tested within
alternative modulated Poisson process models such as the preceding
additive model. We also note that assessment of the dynamics of
individual processes with rather few events is inherently difficult,
and we have found it useful to consider models based on gap times as
well as Poisson models. This is illustrated in Sections~\ref{sec5.1} and
\ref{sec6}.

In choosing a value of $\Delta$, we must rely on background information
that suggests how long a carryover effect might last for the process
under study. Typically $\Delta$ would be fairly small relative to the
average time between events across individuals. The use of specified
durations $\Delta$ for carryover effects is common [e.g., \citet{FarWhi06},
\citet{CooLaw07}, Section 3.8.2], but there
is generally some uncertainty concerning $\Delta$ and it is best to
consider a few separate values. \citet{Xuetal11} have recently
considered uncertainty about $\Delta$ for external carryover effects
but do not discuss estimation of $\Delta$. If we treat $\Delta$ as an
unknown parameter, there is often an estimability issue, because the
profile likelihood for $\Delta$ supports quite a wide range of values.
We examine this in Section~\ref{sec5}, where we find that the asthma
data sets
do not rule out fairly large values of $\Delta$, due partly to the fact
that a carryover effect is partially confounded with heterogeneity. In
addition, as $\Delta$ becomes sufficiently large all events after the
first will lie within the carryover period and an effect $\beta$ as in
(\ref{e2.2}) is confounded with the scale parameter in $\rho_{0}(t)$.

\section{Simulation studies}\label{sec4}

In this section we present the results of simulation studies conducted
to assess when asymptotic normal approximations for parametric score
test statistics are satisfactory, to investigate the tests' power and
to evaluate their robustness with respect to model misspecification.
Because of space limitations, we provide figures and tables for
selected scenarios, and briefly discuss other scenarios. Additional
results are given in \citet{Cgs10} and in the supplementary
material [\citet{CgsLaw}]. We focus on cases
where the null models are homogeneous Poisson processes; results for
nonhomogeneous processes are similar.

We first consider the fixed effects model (\ref{e2.4}) where $\rho_{0} (t;
\gamma) = \gamma$, and the hypothesis of no carryover effect is tested
by using the statistic (\ref{e3.4}). In simulations we took $\gamma=1$, and
generated the $\alpha_{i}$ from the gamma distribution with mean 1 and
variance $\phi= 0.3$. This variance represents a degree of
heterogenity often seen in medical data. Similar results were obtained
for $\phi= 0.6$. The $\alpha_{i}$ were generated once for each
scenario, so that $\alpha_{1}, \ldots, \alpha_{m}$ are fixed across
the repeated samples. To examine the asymptotic normal approximation
for the null distribution of (\ref{e3.4}), we generated 10,000
realizations of
the $m$ homogeneous Poisson processes. In simulations reported below,
scenarios with various combinations of $m$, $\tau$, $\Delta$ were
considered, with $m = 10$, 20, 50, 100 and $\tau_{i} = \tau= 10$.
Results are similar if the $\tau_{i}$ vary, with mean equal to 10. In
practice, we would be interested in small values of $\Delta$, and we
consider $\Delta= 0.0202$, 0.0513 and 0.1054. The inter-event times
satisfy $\Pr(W_{ij} \leq\Delta) = 1 - e^{-\gamma\Delta} = c$ (say),
and with $\gamma= 1$, the preceding values of $\Delta$ give $c =
0.02$, 0.05 and 0.10, respectively. Table~\ref{tab1} presents
empirical\vspace*{1pt} $p$th
quantiles, $\hat{Q}_{p}$, of the 10,000 score statistics $S$ as well as
the estimates $\hat{P}(S > Q_{p})$, where $Q_{p}$ are the standard
normal $p$-quantiles for $p=0.950$, 0.975 and 0.990. The results
indicate that as $m$ increases the standard normal approximation
significantly underestimates right tail probabilities 0.05, 0.025 and
0.01. As the discussion in Section~\ref{sec3.1} indicates, this inaccuracy
reflects the fact that, for fixed $\tau$ and increasing $m$, the $\alpha
_{i}$ are not estimated consistently and (\ref{e3.4}) is not asymptotically
normal. Most applications of the type considered here involve fairly
large $m$ and rather small numbers of events per individual, so we need
an alternative way to get ``honest'' $p$-values. We recommend the use
of simulation to obtain conditional (on $n_{1}, \ldots, n_{m}$)
$p$-values, as described at the end of Section~\ref{sec3.1}.

%
%
\begin{table}\tabcolsep=4pt
\caption{$\hat{Q}_{p}$ is the empirical $p$th quantile of $S$ in
(\protect\ref{e3.4})
computed from 10,000 samples when $\tau= 10$. $\hat{P}(S > Q_{p})$ is
the proportion of the values of $S$ in 10,000 samples which are larger
than the $p$th quantile of a standard normal distribution. The null
model (\protect\ref{e2.4}) has $\rho_{0}(t; \gamma) = 1 $ and $\alpha
_{i} \sim \mbox{gamma}$
($\mbox{mean}=1$, $\mbox{variance} = 0.3$)}
\label{tab1}
\begin{tabular*}{\tablewidth}{@{\extracolsep{\fill}}l r c c c c c c@{}}
\hline
$\bolds{\Delta}$ & \multicolumn{1}{c}{$\bolds{m}$}
& $\bolds{\hat{Q}_{0.950}}$ & $\bolds{\hat{Q}_{0.975}}$
& $\bolds{\hat{Q}_{0.990}}$ &
$\bolds{\hat{P}(S > 1.645)}$ & $\bolds{\hat{P}(S > 1.960)}$
& $\bolds{\hat{P}(S > 2.326)}$\\
\hline
0.0202 & 10 & 1.658 & 2.090 & 2.632 & 0.0515 & 0.0301 & 0.0174\\
& 20 & 1.569 & 1.950 & 2.426 & 0.0433 & 0.0248 & 0.0115\\
& 50 & 1.362 & 1.720 & 2.095 & 0.0292 & 0.0148 & 0.0067\\
& 100 & 1.243 & 1.591 & 1.990 & 0.0226 & 0.0107 & 0.0049\\
[4pt]
0.0513 & 10 & 1.469 & 1.873 & 2.289 & 0.0367 & 0.0206 & 0.0090\\
& 20 & 1.418 & 1.781 & 2.166 & 0.0319 & 0.0168 & 0.0072\\
& 50 & 1.234 & 1.511 & 1.932 & 0.0192 & 0.0096 & 0.0024\\
& 100 & 0.988 & 1.265 & 1.622 & 0.0094 & 0.0045 & 0.0017\\
[4pt]
0.1054 & 10 & 1.361 & 1.685 & 2.139 & 0.0276 & 0.0142 & 0.0074\\
& 20 & 1.242 & 1.599 & 1.981 & 0.0220 & 0.0104 & 0.0045\\
& 50 & 1.013 & 1.365 & 1.703 & 0.0117 & 0.0059 & 0.0026\\
& 100 & 0.751 & 1.047 & 1.417 & 0.0062 & 0.0027 & 0.0008\\
\hline
\end{tabular*}
\end{table}

We next examine the power of (\ref{e3.4}) for tests with size 0.05. In each
scenario described below we used the 10,000 realizations of the $m$
processes represented in Table~\ref{tab1} to estimate $5 \%$ critical
values, so
as to have (approximately) correct type~1 error 0.05. We then estimated
the power of (\ref{e3.4}) by generating $1000$ samples in each
scenario from
the following model:
%
%
\begin{equation}\label{e4.1}\label{EqModelA}
\lambda_{i}(t|H_{i}(t)) =
\alpha_i \exp\bigl\{\beta I\bigl(N_{i}(t^{-})>1\bigr)I\bigl(B_{i}(t) \leq\Delta_0\bigr)\bigr\},
\qquad
i=1, \ldots, m,\hspace*{-28pt}
\end{equation}
where the $\alpha_{i}$ ($i=1, \ldots, m$) are generated from a
gamma distribution with mean 1 and variance $\phi$. We allow $\Delta
_{0}$ to differ from $\Delta$ used in (\ref{e3.4}) in order to check on the
effect of misspecifying $\Delta$. We report here only the results under
the model in (\ref{e4.1}) when $m=20$. Table~\ref{tab2} and further
simulation results
confirm that power increases as $\tau$ and $m$ increase. There is some
loss of power if the assumed value of $\Delta$ is too large (i.e., if
$\Delta> \Delta_{0}$), but little loss if it is too small. We also
examined the effect of using the statistic (\ref{e3.4}) when the $\alpha_{i}$
in (\ref{e2.4}) are actually equal [model (\ref{e4.1}) with $\alpha_{i}
= \alpha$],
so that there is no heterogeneity. There is a slight loss of power
relative to the test based on homogeneous Poisson processes
[\citet{Cgs10}], due to the fact that $m$ values $\alpha_{1},
\ldots, \alpha_{m}$ are estimated instead of a single common value
$\alpha$. However, since failure to recognize heterogeneity can lead to
incorrect rejection of the hypothesis of no carryover effect, the
statistic (\ref{e3.4}) is preferable to the test statistic based on
homogeneous processes.

%
%
\begin{table}
\tablewidth=275pt
\caption{Proportion of times in 1000 samples that test statistic
(\protect\ref{e3.4}) exceeded its 0.05 critical value for the
alternative model (\protect\ref{e4.1})
under various scenarios when $m=20$ and $\phi= 0.3$. Critical values
were estimated from 10,000~simulated~samples}
\label{tab2}
\begin{tabular*}{\tablewidth}{@{\extracolsep{\fill}}l c c c c c@{}}
\hline
& & \multicolumn{2}{c}{$\bolds{\tau= 5}$}& \multicolumn{2}{c@{}}{$\bolds{\tau= 10}$}
\\ [-4pt]
& & \multicolumn{2}{c}{\hrulefill}& \multicolumn{2}{c@{}}{\hrulefill}
\\
$\bolds{\Delta}$ & $\bolds{\Delta_{0}}$
& $\bolds{e^{\beta}=2}$ & $\bolds{e^{\beta}=4}$
& $\bolds{e^{\beta}=2}$ & $\bolds{e^{\beta}=4}$ \\
\hline
& $\frac{2}{3} \Delta$ & 0.174 & 0.675 & 0.290 & 0.908 \\
0.0202 & $\Delta$ & 0.294 & 0.874 & 0.481 & 0.983 \\
& $\frac{4}{3} \Delta$ & 0.298 & 0.889 & 0.473 & 0.988 \\ [6pt]
& $\frac{2}{3} \Delta$ & 0.317 & 0.945 & 0.531 & 0.998 \\
0.0513 & $\Delta$ & 0.543 & 0.994 & 0.821 & 1.000 \\
& $\frac{4}{3} \Delta$ & 0.509 & 0.991 & 0.794 & 1.000 \\ [6pt]
& $\frac{2}{3} \Delta$ & 0.505 & 0.998 & 0.779 & 1.000 \\
0.1054 & $\Delta$ & 0.794 & 1.000 & 0.973 & 1.000 \\
& $\frac{4}{3} \Delta$ & 0.720 & 0.999 & 0.940 & 1.000 \\ \hline
\end{tabular*}
\end{table}

%
%
\begin{table}[b]
\tabcolsep=4pt
\caption{$\hat{Q}_{p}$ is the empirical $p$th quantile of $S$ in
(\protect\ref{e3.6})
computed from 10,000 samples when $m>1$ and $\tau= 10$. $\hat{P}(S >
Q_{p})$ is the proportion of the values of $S$ in 10,000 samples which
are larger than the $p$th quantile of a standard normal distribution.
The null model (\protect\ref{e2.4}) has $\rho_{0}(t; \gamma) = 1 $ and
$\alpha_{i} \sim
\mbox{gamma}$ ($\mbox{mean}=1$, $\mbox{variance} = 0.3$)}\label{tab3}
\begin{tabular*}{\tablewidth}{@{\extracolsep{\fill}}l r c c c c c c@{}}
\hline
$\bolds{\Delta}$ & \multicolumn{1}{c}{$\bolds{m}$}
& $\bolds{\hat{Q}_{0.950}}$ & $\bolds{\hat{Q}_{0.975}}$
& $\bolds{\hat{Q}_{0.990}}$ &
$\bolds{\hat{P}(S > 1.645)}$ & $\bolds{\hat{P}(S > 1.960)}$
& $\bolds{\hat{P}(S > 2.326)}$\\
\hline
0.0202 & 10 & 1.835 & 2.263 & 2.735 & 0.0479 & 0.0303 & 0.0171\\
& 20 & 1.785 & 2.177 & 2.707 & 0.0625 & 0.0370 & 0.0196\\
& 50 & 1.725 & 2.099 & 2.589 & 0.0573 & 0.0326 & 0.0159\\
& 100 & 1.703 & 2.020 & 2.434 & 0.0561 & 0.0284 & 0.0124\\
[4pt]
0.0513 & 10 & 1.779 & 2.179 & 2.656 & 0.0627 & 0.0357 & 0.0192\\
& 20 & 1.694 & 2.080 & 2.458 & 0.0562 & 0.0312 & 0.0146\\
& 50 & 1.691 & 2.027 & 2.404 & 0.0554 & 0.0289 & 0.0120\\
& 100 & 1.665 & 1.997 & 2.361 & 0.0515 & 0.0268 & 0.0111\\
[4pt]
0.1054 & 10 & 1.682 & 2.049 & 2.456 & 0.0534 & 0.0291 & 0.0126\\
& 20 & 1.669 & 2.016 & 2.366 & 0.0523 & 0.0285 & 0.0110\\
& 50 & 1.642 & 2.008 & 2.345 & 0.0497 & 0.0280 & 0.0105\\
& 100 & 1.631 & 1.942 & 2.359 & 0.0479 & 0.0238 & 0.0107\\
\hline
\end{tabular*}
\end{table}

The fixed effects tests are primarily of interest when $m$ is small. We
recommend the random effects tests more generally, and the remaining
discussion concerns them. We first investigated the random effects test
statistic (\ref{e3.6}) for the case where $\rho_{0} (t; \gamma) = \gamma
$ in
(\ref{e2.4}), and the $\alpha_{i}$ were independent gamma random variables
with mean\vadjust{\goodbreak} 1 and variance $\phi= 0.3$. We generated 10,000 replicates
of $m$ homogeneous Poisson processes for $\gamma=1$ and different
combinations of ($\Delta$, $m$, $\tau$) to evaluate the null
distribution and critical values of (\ref{e3.6}). Normal quantile--quantile
plots indicate that the standard normal approximation underestimates
small $p$-values slightly for $m$ less than 50 but is quite good at $m =
100$. Table~\ref{tab3} shows empirical type 1 errors corresponding to normal
errors of 0.01, 0.025 and 0.05 for $\tau= 10$ and $m=10$, 20, 50, 100.
We also generated 1000 samples from versions of model (\ref{e4.1}) to
estimate the power of the test. In each simulation run, we generated a
new set of $\alpha_{i}$ from the gamma distribution with mean 1 and
variance $\phi$. Table~\ref{tab4} shows the results for different
($\Delta_{0}$,
$e^{\beta}$, $m$, $\tau$) combinations and $\phi= 0.3$. The power is
generally high when $e^{\beta} = 3$, with a little decrease when $\Delta
$ is chosen too large. The power values are higher than those for the
fixed effects test in Table~\ref{tab2}, in comparable scenarios. A simulation
study for the power of the statistic (\ref{e3.6}) when $\phi=0.6$ gave similar
results [\citet{Cgs10} and supplementary file,
\citet{CgsLaw}, Table~S.~4].

%
%
\begin{table}
\caption{Proportion of times in 1000 samples that test statistic
(\protect\ref{e3.6}) exceeded its 0.05 critical value for the
alternative model (\protect\ref{e4.1})
under various scenarios when $\phi= 0.3$}\label{tab4}
\label{TaPowerCase3Model2aPhi03}
\begin{tabular*}{\tablewidth}{@{\extracolsep{\fill}}l c c c c c c c@{}}
\hline
& & \multicolumn{2}{c}{$\bolds{m=20}$, $\bolds{\tau= 10}$}
&\multicolumn{2}{c}{$\bolds{m=40}$, $\bolds{\tau= 5}$}
&\multicolumn{2}{c}{$\bolds{m=40}$, $\bolds{\tau= 10}$} \\ [-4pt]
& & \multicolumn{2}{c}{\hrulefill}&
\multicolumn{2}{c}{\hrulefill}& \multicolumn{2}{c@{}}{\hrulefill}
\\
$\bolds{\Delta}$ & $\bolds{\Delta_{0}}$ & $\bolds{e^{\beta}=2}$
& $\bolds{e^{\beta}=3}$
& $\bolds{e^{\beta}=2}$ & $\bolds{e^{\beta}=3}$
& $\bolds{e^{\beta}=2}$ & $\bolds{e^{\beta}=3}$ \\
\hline
& $\frac{2}{3} \Delta$ & 0.282 & 0.693 & 0.316 & 0.692 & 0.493 & 0.936
\\
0.0202 & $\Delta$ & 0.437 & 0.912 & 0.496 & 0.924 & 0.781 & 0.994 \\
& $\frac{4}{3} \Delta$ & 0.460 & 0.886 & 0.498 & 0.914 & 0.776 & 0.994
\\
[6pt]
& $\frac{2}{3} \Delta$ & 0.565 & 0.959 & 0.527 & 0.949 & 0.805 & 0.999
\\
0.0513 & $\Delta$ & 0.828 & 0.997 & 0.809 & 0.998 & 0.979 & 1.000 \\
& $\frac{4}{3} \Delta$ & 0.776 & 0.997 & 0.806 & 0.996 & 0.972 & 1.000
\\
[6pt]
& $\frac{2}{3} \Delta$ & 0.785 & 0.999 & 0.808 & 0.996 & 0.959 & 1.000
\\
0.1054 & $\Delta$ & 0.968 & 1.000 & 0.968 & 1.000 & 1.000 & 1.000 \\
& $\frac{4}{3} \Delta$ & 0.961 & 1.000 & 0.942 & 1.000 & 0.997 & 1.000
\\
\hline
\end{tabular*}
\end{table}

In applications like the ones we consider, the number of individuals
$m$ is usually large but the expected number of events per individual
is small. We next generated 10,000 realizations of $m$ processes under
the model (\ref{e4.1}) with $\phi= 0.6$ and $\beta= 0$, for the cases $m=100$,
200, 500 and $E\{ N_{i}(\tau_{i})\}$ made equal to 1, 2, 5 by
generating $\tau_{i}$ from a uniform distribution over (0.8, 1.2),
(1.6, 2.4) or (4.0, 6.0), respectively. We calculated test statistic
(\ref{e3.6}) for the values of $\Delta= 0.0513$, 0.1054 and 0.2231. The larger
$\Delta$ values reflect features of the data considered in Section
\ref{sec5} and $\phi=0.6$ is between plausible values in the two data
sets there. Normal probability plots of (\ref{e3.6}) and Tables S. 5,
S. 6 and
S. 7 in the supplementary material [\citet{CgsLaw}] show the
standard normal approximation to be quite good except when
$\Delta=0.0513$, $m=100$ and $E\{ N_{i}(\tau_{i})\} = 1$, 2. Once
again, we recommend using simulation (parametric bootstrap) to get
``honest'' $p$-values for such cases. We also conducted a simulation
study to investigate the power of the score statistic (\ref{e3.6}). We used
the 10,000 realizations of the null model discussed above to estimate
5\% critical values. We considered $m =100$, 200, 500 and $\phi= 0.6$,
and generated 1000 realizations of processes with the intensity
function (\ref{e4.1}) for $\exp(\beta) = 1$, 2 and 3. Table~\ref{tab5}
shows power of
(\ref{e3.6}) for the combinations of [$\Delta$, $\Delta_{0}$, $\exp
(\beta)$,
$E\{ N_{i}(\tau_{i})\}$] when $m=200$ (Tables S. 8 and S. 9 in the
supplementary material give the results when $m=100$ and 500, resp.).
Overall, test statistic (\ref{e3.6}) maintains high power in these settings,
and is robust with respect to mild misspecification of $\Delta$.

%
%
\begin{table}
\caption{Proportion of times in 1000 samples that test statistic
(\protect\ref{e3.6}) exceeded its 0.05 critical value for the
alternative model (\protect\ref{e4.1})
under various scenarios when $\phi=0.6$ and $m=200$}\label{tab5}
\begin{tabular*}{\tablewidth}{@{\extracolsep{\fill}}l c c c c c c c@{}}
\hline
& & \multicolumn{2}{c}{$\bolds{E\{ N_{i} (\tau_{i}) \} = 1}$}
&\multicolumn{2}{c}{$\bolds{E\{ N_{i} (\tau_{i}) \} = 2}$}
&\multicolumn{2}{c@{}}{$\bolds{E\{ N_{i} (\tau_{i}) \} = 5}$}\\ [-4pt]
& & \multicolumn{2}{c}{\hrulefill}&
\multicolumn{2}{c}{\hrulefill}& \multicolumn{2}{c@{}}{\hrulefill}
\\
$\bolds{\Delta}$ & $\bolds{\Delta_{0}}$ & $\bolds{e^{\beta}=2}$
& $\bolds{e^{\beta}=3}$ & $\bolds{e^{\beta}=2}$ & $\bolds{e^{\beta}=3}$
& $\bolds{e^{\beta}=2}$ & $\bolds{e^{\beta}=3}$ \\ \hline
& $\frac{2}{3} \Delta$ & 0.585 & 0.975 & 0.858 & 1.000 & 0.995 & 1.000
\\
0.0513 & $\Delta$ & 0.843 & 0.999 & 0.990 & 1.000 & 1.000 & 1.000 \\
& $\frac{4}{3} \Delta$ & 0.809 & 0.998 & 0.985 & 1.000 & 1.000 & 1.000
\\
[6pt]
& $\frac{2}{3} \Delta$ & 0.756 & 1.000 & 0.984 & 1.000 & 1.000 & 1.000
\\
0.1054 & $\Delta$ & 0.952 & 1.000 & 1.000 & 1.000 & 1.000 & 1.000 \\
& $\frac{4}{3} \Delta$ & 0.873 & 1.000 & 0.997 & 1.000 & 1.000 & 1.000
\\
[6pt]
& $\frac{2}{3} \Delta$ & 0.803 & 1.000 & 0.999 & 1.000 & 1.000 & 1.000
\\
0.2231 & $\Delta$ & 0.951 & 1.000 & 1.000 & 1.000 & 1.000 & 1.000 \\
& $\frac{4}{3} \Delta$ & 0.842 & 0.999 & 0.999 & 1.000 & 1.000 & 1.000
\\
\hline
\end{tabular*}
\end{table}

Finally, simulation studies were conducted to examine the performance
of the test statistic (\ref{e3.6}) when the assumption that the $\alpha_{i}$
have a gamma distribution is not true. To do that, we generated the
$\alpha_{i}$ from a lognormal distribution with mean 1 and variance
$\phi$. We then generated 1000 realizations of $m$ processes when
$\Delta= 0.0202$ and $e^{\beta} = 1$, 2, 3, 4, and calculated the
proportion of the time that (\ref{e3.6}) exceeded the 0.05 critical value.
Results are given in Supplementary Table S. 10, for scenarios with $\tau
=10$ and $m=20$, 40. The column $e^{\beta} = 1$ shows the empirical
type 1 errors based on the 1000 samples; they are close to the nominal
significance level 0.05. In addition, (\ref{e3.6}) maintains high power in
this case, and we conclude that mild misspecification of the
distribution of random effects is not a problem; this agrees with
similar results for estimation of rate functions in mixed Poisson
processes without carryover effects [\citet{Law87}].

\section{Applications}\label{sec5}

\subsection{Recurrent asthma attacks in children (I)}\label{sec5.1}

Duchateau et~al. (\citeyear{Ducetal03}) discussed data from a
prevention trial in
infants with a high risk of asthma, but without a prior attack. The
subjects were 6 months of age on entry to the study. The follow-up
period for each subject was approximately 18 months, and started after
random allocation to a placebo control group or an active drug
treatment group. The main aim of the study was to assess the effect of
the drug on the occurrence of asthma attacks, but an interesting
secondary question was whether the occurrence of an event (asthma
attack) influences the future event rate. There were 483 asthma attacks
among 119 children in the control group and 336 asthma attacks among
113 children in the treatment group, during the 18 month follow-up.

The Nelson--Aalen estimates of the mean function
[\citet{CooLaw07}, Section 3.4] for each treatment group are close
to linear but that does not in itself show that the possibly
heterogeneous individual rate functions are constant. Therefore, we
fitted models (\ref{e2.4}) in which $\rho_{0}(t)$ took the power law form
$\gamma_{1} \gamma_{2} t^{\gamma_{2}-1}$. We found no evidence against
the constancy of $\rho_{0}(t)$, and so the following details are based
on constant rates which may vary across individuals. A caveat
concerning the data is that \citet{Ducetal03} do not provide the
trial entry dates for each subject, so it is not possible to assess
whether there might be a seasonal effect. However, for the second
asthma data set considered in Section~\ref{sec5.2}, such information
was available and no seasonal effect was seen. An asthma attack lasts
an average of 6--7 days, and a patient is not considered at-risk for a
new attack over that time. The at-risk indicator $Y_{i}(t)$ takes value
1 if subject $i$ is at risk of an asthma attack at time $t$, and the
intensity model for subject $i$ that we consider is therefore
%
%
\begin{equation}\label{e5.1}\label{Eqmodelasthma}
\lambda_{i} ( t| \mathcal{H}_{i} ( t ) ) =
Y_{i}(t) \alpha_{i} \gamma\exp\{ \beta Z_{i}(t)\},\qquad t \geq0,
\end{equation}
where $Z_{i}(t) = I\{N_{i}(t^{-})>0\} I\{ B_{i}(t) \leq\Delta\}$, and
$B_{i}(t)$ is the time since the subject $i$ started their current
at-risk period.

We will consider the treatment and control groups separately. To allow
for heterogeneity, we use the tests of Section~\ref{sec3} with the random
effects model (\ref{e5.1}), where $\alpha_{i} \sim \operatorname{Gamma}(1,\phi)$, for testing
$H_{0}\dvtx\beta= 0$. Results obtained by fitting models with a range of
values for $\Delta$ are shown in Table~\ref{tab6}; to conserve space, standard
errors for estimates $\hat{\phi}$ are not given, but in every model
heterogeneity ($\phi> 0$) is strongly significant.

%
%
\begin{table}
\tabcolsep=4pt
\begin{center}
\caption{The results of the no carryover test based on (\protect\ref
{e3.6}) for
various $\Delta$ values.\break $\operatorname{Exp}(\Delta)$ is the second term on the
right-hand
side of (\protect\ref{e3.7}). $Z^{2}$ is\break the square of $\hat{\beta}/
\mathit{se}(\hat{\beta})$,
and $\ell_{\mathrm{max}} = \ell(\hat{\gamma}, \hat{\beta}, \hat{\phi})$}\label{tab6}
\label{Taparameters1}
\begin{tabular*}{\tablewidth}{@{\extracolsep{\fill}}l r d{3.0} d{3.3} c c c c c d{5.2}@{}}
\hline
\textbf{Group} & $\bolds{\Delta}$ & \multicolumn{1}{c}{$\bolds{\operatorname{Obs}(\Delta)}$}
& \multicolumn{1}{c}{$\bolds{\operatorname{Exp}(\Delta)}$} & $\bolds{\hat{\gamma}}$ &
$\bolds{\hat{\beta}}$ & $\bolds{\hat{\phi}}$ & $\bolds{S^{2}}$ & $\bolds{Z^{2}}$ &
\multicolumn{1}{c@{}}{$\bolds{\ell_{\mathrm{max}}}$} \\
\hline
\textit{Treatment}
& 7 & 40 & 22.858 & 0.006 & 0.681 & 0.476 & 14.900 & 14.314 & -2009.41
\\
& 14 & 76 & 40.464 & 0.005 & 0.904 & 0.388 & 40.513 & 33.338 &
-1998.52 \\
& 28 & 119 & 67.099 & 0.005 & 1.017 & 0.305 & 61.968 & 59.360 &
-1988.08 \\
& 42 & 143 & 86.213 & 0.004 & 1.015 & 0.284 & 65.206 & 62.880 &
-1985.84 \\
& 56 & 162 & 101.774 & 0.004 & 1.029 & 0.270 & 68.857 & 66.694 &
-1983.75 \\
& 70 & 171 & 114.660 & 0.004 & 0.942 & 0.288 & 57.882 & 56.791 &
-1988.47 \\
[4pt]
\textit{Control}
& 7 & 68 & 47.173 & 0.008 & 0.486 & 0.521 & 11.751 & 11.551 & -2726.18
\\
& 14 & 121 & 80.302 & 0.007 & 0.637 & 0.455 & 29.921 & 29.142 &
-2717.95 \\
& 28 & 185 & 130.457 & 0.007 & 0.678 & 0.399 & 40.284 & 39.411 &
-2712.53 \\
& 42 & 227 & 167.050 & 0.006 & 0.699 & 0.373 & 43.944 & 43.485 &
-2710.27 \\
& 56 & 260 & 195.336 & 0.006 & 0.745 & 0.350 & 49.393 & 48.478 &
-2707.26 \\
& 70 & 272 & 218.287 & 0.006 & 0.622 & 0.383 & 33.698 & 33.169 &
-2714.75 \\
\hline
\end{tabular*}
\end{center}
\end{table}

Table~\ref{tab6} gives, for each value of $\Delta$, the estimates of
$\gamma$,
$\beta$ and $\phi$ in model (\ref{e5.1}), along with the squared score
statistic $S^{2}$ [with\vspace*{1pt} $S$ given by (\ref{e3.6})] and a corresponding Wald
statistic for testing $\beta= 0$, defined as $Z^{2} = \hat{\beta}^{2}
/ \widehat{\operatorname{Var}} (\hat{\beta})$. The models were fitted using R function
\texttt{nlm}, which automatically provides variance estimates via
numerical differentiation. The score statistic $S$ is more easily
obtained since only restricted estimates $\tilde{\gamma}$ and $\tilde
{\phi}$ are needed, but computational differences are unimportant here.
The two statistics agree closely and strongly contradict the hypothesis
($\beta= 0$) of no carryover effect for every value of $\Delta$ shown.
The $p$-values obtained from $\chi^{2}_{(1)}$ approximations for
$S^{2}$ and $Z^{2}$ are virtually zero. As a check on this we also
obtained $p$-values for $S^{2}$ by simulating 1000 samples under the
null model with parameter values $\tilde{\gamma}$, $\tilde{\phi}$. For
each value of $\Delta$, there were no samples out of the 1000
generated in which $S^{2}$ exceeded its observed value in the data set.
We also show observed and expected numbers [$\operatorname{Obs}(\Delta)$, $\operatorname{Exp}(\Delta
)$] of events in carryover periods, assuming no carryover effect; these
are given in (\ref{e3.7}). This provides a nice summary of the excess events
observed within time $\Delta$ of a preceding event.

Table~\ref{tab6} indicates that a wide range of values for $\Delta$ is
plausible. We show the maximum values $\ell_{\mathrm{max}}$ of the log
likelihood for each model, and see that the value of $\Delta$ (among
those shown) best supported by the data is $\Delta= 56$ days in both
the treatment and control groups. It is also seen in Table~\ref{tab6} that
estimates $\hat{\beta}$ and $\hat{\phi}$ are negatively correlated, as
our discussion in Section~\ref{sec2} suggests. As $\Delta$ increases further
beyond 70 days, the values of $\ell_{\mathrm{max}}$ continue to decrease, and
$\Delta= 100$ days still gives values that are about the same as
$\Delta= 14$ days. The values of $\tilde{\gamma}$ in the treatment and
control groups, respectively, are 0.00608 and 0.00822, indicating an
average of about one event every 165 days per subject in the treatment
group, and one event every 122 days for the control group. The evidence
indicates that events tend to occur closer to the previous event more
often than is expected under a homogeneous Poisson process.

Our results can also be interpreted as indicating that the gap times
between successive asthma attacks do not follow exponential
distributions for individual subjects. \citet{Ducetal03} fitted
models in which gap times are assumed to be independent Weibull random
variables within individuals, and heterogeneity is incorporated through
individual-level gamma-distributed random effects. They found strong
evidence of a decreasing hazard function for gap times, which is
consistent with a carryover effect. The Duchateau et~al. model has $p =
4$ parameters and ours has three, but AIC values ($-2 \ell_{\mathrm{max}} + 2 p
$) are very close. For example, in the control group we find the AIC
for (\ref{e5.1}) with $\Delta= 14$ days as 5441.9 ($p = 3$) and the AIC for
the Duchateau et~al. model as 5437.6 ($p = 4$). Smaller AIC values are
obtained for models (\ref{e5.1}) with larger values of $\Delta$; for example,
when $\Delta= 56$ the AIC for the control group is 5420.5, the smallest
for the models considered here.

Thus, all models indicate that the probability of a new asthma attack
is highest soon after a preceding attack, and then decreases. Whether a
delayed renewal process or a modulated Poisson process best describe
the situation is not clear, nor whether there is a carryover effect of
limited duration or a smooth decreasing hazard function for gap times.
Without additional information concerning the asthma attacks and their
treatment, we also cannot know the basis of the perceived effect.

\subsection{Recurrent asthma attacks in children (II)}\label{sec5.2}

We now briefly consider the randomized trial on the effects of 200
versus 400 $\mu$g per day of fluticasone propionate (FP) in preventing
asthma attacks in children, mentioned in Section~\ref{sec1}.
\citet{Veretal03} describe the study in detail, and the data have
been reanalyzed
by \citet{CooLaw07}, Section 5.5.2. None of the previous
analyses has looked at the interesting secondary issue of whether there
is any indication of a carryover effect; we consider this here.

Earlier analyses showed that age and predicted expiratory flow (PEF) at
enrollment had some predictive power for asthma exacerbations and we
included them in our models. Seasonal effects and covariates such as
sex and weight were examined but were not found significant and are
excluded here. We ran analyses based on the modulated Poisson model
(\ref{e2.4}) with gamma random effects and different values of $\Delta$ for
the duration of carryover. In the interest of brevity we focus here on
the FP200 group, which had 267 subjects. About one-third had
approximately 3 months follow-up, with two-thirds followed for
approximately 12 months. Cook and Lawless [(\citeyear{CooLaw07}), page
195] show the numbers of asthma attacks per subject; there were a total
of 359 in the FP200 group. As an illustration of the semiparametric
approach we used the Andersen--Gill version of (\ref{e2.4}) with additional
covariates, so no parametric assumption concerning $\rho_{0}(t)$ was
made. According to the protocol for the trial, an exacerbation was
counted only if it was not within 10 days of the start of a previous
exacerbation, so the at-risk indicator $Y_{i}(t)$ introduced in (\ref{e5.1})
is defined so that $Y_{i}(t)$ equals 1 if and only if subject $i$ is
not within 10 days of a preceding exacerbation.

%
%
\begin{table}
\tablewidth=197pt
\caption{Estimation results for Andersen--Gill models
(\protect\ref{e2.4}) with gamma
random effects, fitted to FP200 asthma trial~data}\label{tab7}
\label{Taasthma2}
\begin{tabular*}{\tablewidth}{@{\extracolsep{\fill}}l c c c d{2.2} d{5.2}@{}}
\hline
$\bolds{\Delta}$\textbf{\tabnoteref{ta}} & $\bolds{\hat{\beta}}$
& $\bolds{\mathit{se}(\hat{\beta
})}$ & $\bolds{\hat{\phi}}$ & \multicolumn{1}{c}{$\bolds{Z^{2}}$\textbf{\tabnoteref{tb}}} &
\multicolumn{1}{c@{}}{$\bolds{\ell_{\mathrm{max}}}$} \\
\hline
\hphantom{0}7 & 0.206 & 0.185 & 1.58 & 1.24 & -1800.28 \\
14 & 0.394 & 0.144 & 1.46 & 2.49 & -1797.56 \\
28 & 0.241 & 0.133 & 1.47 & 3.28 & -1799.43 \\
42 & 0.426 & 0.130 & 1.27 & 10.34 & -1796.77 \\
56 & 0.487 & 0.132 & 1.18 & 13.61 & -1796.00 \\
70 & 0.462 & 0.134 & 1.19 & 11.89 & -1796.89 \\
84 & 0.419 & 0.136 & 1.21 & 9.49 & -1797.78 \\
\hline
\end{tabular*}
\tabnotetext[a]{ta}{$\Delta$ is in days.}
\tabnotetext[b]{tb}{$Z^{2} = \hat{\beta}^{2} / \mathit{se}(\hat{\beta
})^{2}$.}\vspace*{-3pt}
\end{table}

Table~\ref{tab7} shows results for models fitted with various values of
carryover duration~$\Delta$; models were fitted using R function
\texttt{coxph}. As in the preceding example, there is strong evidence
against the hypothesis of no carryover effect ($\beta= 0$), but a wide
range of values for $\Delta$ is supported by the data. The best
supported value is about 56 days (8 weeks), as in the study in Section~\ref{sec5.1}.
The average rate of events per subject in these data is
about 1.8 per year, or about one asthma attack every 29 weeks.
Therefore, there is once again an indication that the risk of a new
attack is higher soon after a previous attack. As in the preceding
case, there is also strong evidence of heterogeneity across subjects.
This information, in conjunction with background medical information,
may suggest that modifications to the prevention or treatment of
attacks be considered.\vspace*{-3pt}

\section{Concluding remarks}\label{sec6}

We have considered modulated Poisson process models and tests for
carryover effects, allowing for time trends and heterogeneity across
processes. The random effects models and tests are recommended for
general use; the tests have better power and are better approximated by
asymptotic normal theory, especially when $m$ is large. Fixed or
time-varying covariates can be incorporated into our approach, as
illustrated in Section~\ref{sec5.2}.

It can be hard to deconstruct the dynamics of event occurrence when
there are few events for most individuals, and the examination of
alternative models is important. An alternative approach that is useful
is to examine the distribution of ``gap'' times between successive
events. The presence of a carryover effect is suggested by the density
or hazard function for the gap times having substantial mass near zero.
Such models do not produce definitions or tests for a carryover effect
or handle time trends as readily as the models in Section~\ref{sec3}. However,
examination of gap time models as in Section~\ref{sec5.1} is often
helpful, and
in the absence of covariates, nonparametric estimates of hazard or
density functions for gap times are useful. As an additional
illustration, we consider data on children with hydrocephalus, who have\vadjust{\goodbreak}
shunts inserted to drain excess cerebrospinal fluid. In the study
mentioned in Section~\ref{sec1} [\citet{Tuletal00}], data on 839
children who
had initial shunts inserted during the years 1989--1996 at one Canadian
hospital were analyzed. Such shunts can ``fail'' due to blockages,
infections and other conditions, necessitating full or partial
replacement of the shunt. The data in question were analyzed previously
by \citet{Lawetal01} and \citet{CooLaw07},
Section 5.4.2.
Gap time models are a natural approach in this case: the occurrence of
a failure results in a new shunt, and it makes sense to examine the
lifetime of each subsequent shunt. The previous analyses were based on
Cox models fitted to the survival times of successive shunts, and they
showed that there were several important covariates, including the
cause of a child's hydrocephalus and the age of the child at the time a
shunt was (surgically) inserted. They also showed a tendency for second
or third shunts to fail sooner than initial shunts. Plots of estimated
baseline cumulative hazard functions $\tilde{H}_{0j} (t)$ for shunts
$j=1, 2, \ldots$ [e.g., \citet{CooLaw07}, Figure 5.9]
suggested that the risk of failure was high soon after a new shunt was
inserted, but this was not examined further. Table~\ref{tab8} shows a
discretized estimate of the baseline hazard functions $h_{02}(w)$ and
$h_{03}(w)$ for second and third shunts for a model involving
adjustment for important covariates and additional allowance for
heterogeneity. The covariates are coded for the two models such that
the baseline hazard functions $h_{02}(w)$ and $h_{03}(w)$ represent the
same vector of covariate values. The estimates are piecewise-constant,
with $\tilde{h}_{0j}(w) = [\tilde{H}_{0j}(a_{j}) - \tilde
{H}_{0j}(a_{j-1})]/(a_{j}-a_{j-1})$ for $a_{j-1} < w \leq a_{j}$ and
$a_{j} = 0, 60, 120, \ldots$ (days) for $j=0, 1, 2, \ldots\,$.
%
%
\begin{table}
\tablewidth=262pt
\caption{Estimates of baseline cumulative hazard and
piecewise-constant hazard functions for cerebrospinal fluid
shunts}\label{tab8}
\label{Taasthma2}
\begin{tabular*}{\tablewidth}{@{\extracolsep{\fill}}l d{1.5} c d{1.5} c@{}}
\hline
& \multicolumn{2}{c}{\textbf{Second shunts}} & \multicolumn{2}{c@{}}{\textbf{Third
shunts}} \\
[-4pt]
& \multicolumn{2}{c}{\hrulefill}& \multicolumn{2}{c@{}}{\hrulefill}
\\
$\bolds{a}$ & \multicolumn{1}{c}{$\bolds{\hat{H}_{02}(a)}$}
& \multicolumn{1}{c}{$\bolds{\hat{h}_{02}(a)}$\textbf{\tabnoteref{tc}}}
& \multicolumn{1}{c}{$\bolds{\hat{H}_{03}(a)}$} &
\multicolumn{1}{c@{}}{$\bolds{\hat{h}_{03}(a)}$\textbf{\tabnoteref{tc}}} \\
\hline
\hphantom{00}0 & 0 & &  0 & \\
\hphantom{0}60 &0.19263 &0.00321 & 0.30316 &0.00505 \\
120 &0.26584 &0.00122 & 0.39280 &0.00149 \\
180 &0.29106 &0.00042 & 0.44324 &0.00084 \\
240 &0.32343 &0.00054 & 0.47100 &0.00046 \\
300 &0.35529 &0.00053 & 0.48560 &0.00024 \\
360 &0.38951 &0.00057 & 0.51582 &0.00050 \\
\hline
\end{tabular*}
\tabnotetext[a]{tc}{$\hat
{h}_{0j}(a) = [\hat{H}_{0j}(a) - \hat{H}_{0j}(a-60)]/60$, $j =
2$, 3.}
\end{table}
It is seen that the hazard functions are sharply decreasing. The time
to failure of the initial shunt also shows a decreasing hazard
function, but with overall smaller values. This indicates the risk of
shunt failure is highest soon after it is inserted, and one explanation
is that problems leading to a shunt failure may in some cases persist
and create problems for the new shunt.\vadjust{\goodbreak}

Finally, in many settings events of different types may occur. For
example, that is the case with shunt failures, which can be due to
obstruction, infection or other causes. In this context we can specify
separate covariates to represent carryover effects related to the
different event types. This is readily done in either the modulated
Poisson process framework or the gap time framework. Table~\ref{tab8}
is from a
combined-causes analysis of the shunt failures, but separate causes
could be considered similarly.

%
\begin{appendix}\label{app}

\section{Andersen--Gill model}\label{sec7}\label{appA}

For the modulated Andersen--Gill model (\ref{e2.2}) for recurrent
events, the
Cox partial likelihood function for $\beta$ gives the score function
[\citet{CooLaw07}, page 71]
%
%
\begin{equation}\label{eA.1}
U_{\beta}(\beta) = \sum_{i=1}^{m} \biggl\{ \int_{0}^{\tau_{i}} Z_{i}(t)
\biggl[ dN_{i}(t) - \frac{d \bar{N}_{\cdot}(t)
e^{\beta^{\prime}Z_{i}(t)}} {\sum_{l=1}^{m}Y_{l}(t)
e^{\beta^{\prime}Z_{l}(t)}} \biggr] \biggr\},
\end{equation}
where $dN_{i}(t) = I$ (process $i$ has an event at time $t$),
$Y_{l}(t)= I(\tau_{l} \geq t)$, and $d \bar{N}_{\cdot}(t) = \sum_{l=1}^{m}
Y_{l}(t) \,dN_{l}(t)$. The score statistic at $\beta= 0$ is
%
%
\begin{equation}\label{eA.2}
U_{\beta}(0) = \sum_{i=1}^{m} \biggl\{ \int_{0}^{\tau_{i}} Z_{i}(t) [
dN_{i}(t) - \tilde{\rho}_{0}(t) \,dt ] \biggr\},
\end{equation}
where, taking liberties with notation,
%
%
\begin{equation}\label{eA.3}
\tilde{\rho}_{0} (t) \,dt =
\frac{d \bar{N}_{\cdot}(t)}
{\sum_{l=1}^{m} Y_{l}(t)}
\end{equation}
is the estimated baseline rate function at time $t$. Thus, (\ref{eA.1})
can be
rewritten in ``\textit{Observed}--\textit{Expected}'' form as
%
%
\begin{equation}\label{eA.4}
U_{\beta}(0) =
\sum_{i=1}^{m} \sum_{j=1}^{n_{i}} Z_{i}(t_{ij})-
\sum_{r=1}^{R} Z_{\cdot}(t^{\ast}_{r})\,
\frac{d\bar{N}_{\cdot}(t^{\ast}_{r})}
{Y_{\cdot}(t^{\ast}_{r})},
\end{equation}
where $t_{1}^{\ast}, \ldots, t_{R}^{\ast}$ are the distinct event
times across all processes, and $Z_{\cdot}(t) = \sum_{i=1}^{m} Z_{i}(t)$,
$Y_{\cdot}(t) = \sum_{i=1}^{m} Y_{i}(t)$, and $d \bar{N}_{\cdot} (t)$ is
defined following (\ref{eA.1}). This approach can be used if there is no
evidence of heterogeneity across individuals. Usually this is not the
case and then we should use the approach described at the end of
Appendix~\ref{appB}.

\section{Score statistics for gamma random effects~models}\label
{sec8}\label{appB}

We consider here the score statistic (\ref{e3.6}) arising from the log
likelihood (\ref{e3.5}). The numerator is easily shown to be
%
%
\begin{eqnarray}\label{eB.1}
U_{\beta} ( \tilde{\gamma}, 0, \tilde{\phi} ) & = &
\biggl(
\frac{\partial\ell(\gamma, 0, \phi)}{\partial\beta}
\biggr)_{(\tilde{\gamma}, 0, \tilde{\phi})} \nonumber\\[-8pt]\\[-8pt]
& = &
\sum^{m}_{i=1}
\Biggl\{
\sum^{n_{i}}_{j=1} Z_{i} (t_{ij} ) - \frac{( 1 + \tilde{\phi} n_{i} )\,
\partial R_{i} ( \tilde{\gamma}, 0 ) / \partial\beta}
{1 + \tilde{\phi} R_{i} (\tilde{\gamma}, 0)}
\Biggr\},\nonumber
\end{eqnarray}
where $R_{i} (\gamma, \beta)$ is given by (\ref{e3.2}). A variance estimate
for $U_{\beta} ( \tilde{\gamma}, 0, \tilde{\phi})$ under $H_{0}$ is
given by asymptotic theory for counting processes in the case where $m
\rightarrow\infty$ [\citet{Andetal93}, Chapter 6,
\citet{Pen98}]. This takes the standard form
%
%
\begin{equation}\label{eB.2}
\widehat{\operatorname{Var}}
\{
U_{\beta} ( \tilde{\gamma}, 0, \tilde{\phi} )
\}
=
\tilde{I}_{\beta\beta} -
\pmatrix{
\tilde{I}_{\beta\gamma} & \tilde{I}_{\beta\phi}}
\pmatrix{
\tilde{I}_{\gamma\gamma} & \tilde{I}_{\gamma\phi} \cr
\tilde{I}_{\phi\gamma} & \tilde{I}_{\gamma\gamma}}^{-1}
\pmatrix{
\tilde{I}_{\gamma\beta} \cr
\tilde{I}_{\phi\beta}}.
\end{equation}

The $2 \times2$ matrix in (\ref{eB.2}) is the inverse of the negative Hessian
matrix for the log likelihood $\ell(\gamma, 0, \phi)$ evaluated at
$\tilde{\gamma}$, $\tilde{\phi}$, and the terms $\tilde{I}_{\beta\beta
}$, $\tilde{I}_{\beta\gamma}$ and $\tilde{I}_{\beta\phi}$ are based
on the following, evaluated at $\tilde{\gamma}, \beta= 0, \tilde{\phi}$:
\begin{eqnarray*}
I_{\beta\beta} & = & \frac{- \partial^{2} \ell(\gamma, \beta, \phi
)}{\partial\beta^{2}} \\
&=& \sum^{m}_{i = 1} ( n_{i} + \phi^{- 1} ) \biggl\{
\frac{[ \phi^{- 1} + R_{i} (\gamma, \beta) ] [ \partial R_{i} /
\partial\beta] - [ \partial R_{i} / \partial\beta]^{2}}{[ \phi
^{-1} + R_{i} (\gamma, \beta) ]^{2}} \biggr\},\\
I_{\beta\gamma} & = & \frac{- \partial^{2} \ell(\gamma, \beta, \phi
)}{\partial\beta\,\partial\gamma^{\prime}} \\
&=& \sum^{m}_{i = 1} ( n_{i}
+ \phi^{- 1} ) \biggl\{ \frac{[ \phi^{- 1} + R_{i} (\gamma, \beta) ] [
\partial^{2} R_{i} / \partial\beta\,\partial\gamma^{\prime} ] - [
\partial R_{i} / \partial\beta] [ \partial R_{i} / \partial\gamma
^{\prime} ]}{[ \phi^{-1} + R_{i} (\gamma, \beta) ]^{2}} \biggr\},\\
I_{\beta\phi} & = &
\frac{- \partial^{2} \ell(\gamma, \beta, \phi)}
{\partial\beta\,\partial\phi} = \sum^{m}_{i = 1}
\biggl\{
\frac{( \partial R_{i} / \partial\beta)
[ n_{i} - R_{i} (\gamma, \beta) ] }
{[
1 + \phi R_{i} (\gamma, \beta)
]^{2}}
\biggr\}.
\end{eqnarray*}

The Andersen--Gill model of Appendix~\ref{appA} with added frailty can
be handled
by the R/S-Plus Cox model function \texttt{coxph}. This implementation
returns an estimate $\hat{\beta}$ and standard error, as well as a
maximum likelihood value, so that a likelihood ratio or Wald test of
$\beta= 0$ can be used. A score statistic analogous to (\ref{eB.1}) for the
Cox model is
\[
U_{\beta}(0) =
\sum_{i=1}^{m}
\Biggl\{
\sum_{j=1}^{n_{i}} Z_{i} (t_{ij}) -
\frac{(1+\tilde{\phi} n_{i})
\sum_{t_{r}^{\ast} \leq\tau_{i}} Z_{i}(t_{r}^{\ast})
({dN_{\cdot}(t_{r}^{\ast})}/{Y_{\cdot}(t_{r}^{\ast})})}
{1+\tilde{\phi} \sum_{t_{r}^{\ast} \leq\tau_{i}}
({dN_{\cdot}(t_{r}^{\ast})}/{Y_{\cdot}(t_{r}^{\ast})})}
\Biggr\},
\]
where the $t_{r}^{\ast}$, $dN_{\cdot}(t_{r}^{\ast})$ and $Y_{\cdot
}(t_{r}^{\ast
})$ are as defined in Appendix~\ref{appA}. This statistic has the form
``\textit{Observed}--\textit{Expected};'' the function \texttt{coxph}
does not give it
as output so some additional coding is required.
\end{appendix}

\begin{supplement}
\stitle{Additional simulation results}
\slink[doi]{10.1214/12-AOAS560SUPP} 
\slink[url]{http://lib.stat.cmu.edu/aoas/560/supplement.pdf}
\sdatatype{.pdf}
\sdescription{The supplementary file contains detailed simulation
results to support the discussion in Sections~\ref{sec2} and \ref
{sec4}. Each
simulation study in the supplementary file has its own description and
title.}
\end{supplement}

%

\printaddresses

\end{document}